\documentclass{IEEEtran4PSCC}

\usepackage{comment}
\usepackage{bm}
\usepackage{color}
\usepackage{booktabs}
\usepackage{hhline}

\ifCLASSINFOpdf
   \usepackage[pdftex]{graphicx}
\else
   \usepackage[dvips]{graphicx}
\fi
\usepackage{amstext,amssymb,amsbsy,amsmath}

\usepackage{lipsum}
\makeatletter
\let\old@ps@headings\ps@headings
\let\old@ps@IEEEtitlepagestyle\ps@IEEEtitlepagestyle
\def\psccfooter#1{%
    \def\ps@headings{%
        \old@ps@headings%
        \def\@oddfoot{\strut\hfill#1\hfill\strut}%
        \def\@evenfoot{\strut\hfill#1\hfill\strut}%
    }%
    \def\ps@IEEEtitlepagestyle{%
        \old@ps@IEEEtitlepagestyle%
        \def\@oddfoot{\strut\hfill#1\hfill\strut}%
        \def\@evenfoot{\strut\hfill#1\hfill\strut}%
    }%
    \ps@headings%
}
\makeatother

\psccfooter{%
        \parbox{\textwidth}{\hrulefill \\ \small{23rd Power Systems Computation Conference} \hfill \begin{minipage}{0.2\textwidth}\centering \vspace*{4pt} \includegraphics[scale=0.06]{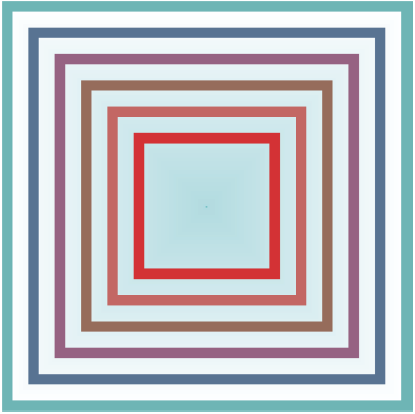}\\\small{PSCC 2024} \end{minipage} \hfill \small{Paris, France --- June, 2024}}%
}
\renewcommand{\baselinestretch}{0.9825}
\begin{document}
%
\title{Privacy-Preserving Distributed Market Mechanism for Active Distribution Networks}


\author{
\IEEEauthorblockN{Matthias Franke\IEEEauthorrefmark{1}, Ognjen Stanojev\IEEEauthorrefmark{1}, Lesia Mitridati\IEEEauthorrefmark{2}, Gabriela Hug\IEEEauthorrefmark{1}}
\IEEEauthorblockA{\IEEEauthorrefmark{1}EEH - Power Systems Laboratory,
ETH Z\"urich,
Z\"urich, Switzerland\\
\IEEEauthorblockA{\IEEEauthorrefmark{2} Center for Electric Power and Energy, Technical University of Denmark, Kgs. Lyngby, Denmark} %
\{mfranke, ognjens, ghug\}@ethz.ch, lemitri@dtu.dk}
}


\maketitle


\begin{abstract}
Amidst the worldwide efforts to decarbonize power networks, Local Electricity Markets (LEMs) in distribution networks are gaining importance due to the increased adoption of renewable energy sources and prosumers. Considering that LEMs involve data exchange among independent entities, privacy and cybersecurity are some of the main practical challenges in LEM design. This paper proposes a secure market protocol using innovations from distributed optimization and Secure MultiParty Computation (SMPC). The considered LEM is formulated as an uncertainty-aware joint market for energy and reserves with affine balancing policies. To achieve scalability and enable the use of SMPC, market clearing is solved using the Consensus ADMM algorithm. Subsequently, the data exchange among participants via ADMM iterations is protected using the Shamir secret-sharing scheme to ensure privacy. The market protocol is further reinforced by a secure and verifiable settlement process that uses SMPC and ElGamal commitments to verify market quantities and by a secure recovery scheme for missing network measurements. Finally, the feasibility and performance of the proposed LEM are evaluated on a 15-bus test network.
\end{abstract}

\vspace{-0.25cm}
\section{Introduction} \label{section:1}
With the goal to mitigate climate change, an increasing effort is made to decarbonize power networks. This effort primarily involves the increased adoption of Distributed Energy Resources (DERs) in distribution grids and empowering households to become \textit{prosumers} that can contribute to decarbonization~\cite{Charbonnier2022}. 
Despite these advancements, current electricity markets still operate in a hierarchical, top-down manner, without adapting their market structures to the emergence of DERs \cite{Sousa2019}.
Furthermore, technical challenges arise in the operation of distribution networks due to high bi-directional power flows and voltage fluctuations \cite{Kim2020,Stanojev2022MultipleNetworks}. These challenges suggest the development of Local Electricity Markets (LEMs) in which small-scale producers and consumers can transact electricity in a decentralized fashion while respecting the physical limits of the distribution network. Nevertheless, such a collaborative principle involves data exchange among independent entities, therefore rising privacy and cybersecurity concerns. 
We propose an uncertainty-aware LEM that is operated in a secure and distributed manner using innovations from optimization and theoretical cryptography.

There has been a wide range of research into innovating LEMs that promise to safely and efficiently operate future networks and markets with DERs \cite{Charbonnier2022}. The three main approaches for such LEMs are~\cite{Sousa2019}: (i) pure Peer-to-Peer markets, whose fully decentralized approach offers customers high flexibility at the cost of lacking convergence and safety guarantees, (ii) community markets with an operator efficiently managing trading both locally and with the upper-level grid, however, resulting in computational scaling issues, and (iii) hybrid methods that use a bilevel approach to leverage the advantages of the former two. Of key interest for this paper are the community markets, which are proven to successfully operate LEMs with DERs in the existing literature \cite{Ratha2019, Mieth2019}. However, these markets have encountered challenges due to the use of oversimplified network models \cite{Ratha2019}, the absence of DER uncertainty management \cite{Sousa2019}, and the lack of effective coordination with the upper-level grid \cite{Mieth2019}. Additionally, most works in this field focus on theoretical explorations of such markets, and there is thus a lack of research into how to provide cyber security and resilience to measurement failure for the actual implementation of LEMs.

Efforts to overcome these gaps have been made, although they have not been consolidated yet. 
For instance, improved models of distribution networks together with techniques to capture the stochastic nature of DERs are available \cite{Zugno2013}, such as the LinDistFlow formulation \cite{Molzahn2017, Zhang2015} that achieves high accuracy through the utilization of linear and convex constraints.
This formulation overcomes shortcomings of the DC power flow formulations for distribution networks \cite{Molzahn2017} and is further highly suitable for adding chance constraints to capture the uncertainty of DERs within a market model \cite{Mieth2019}. Of particular interest to this paper is the usage of chance constraints to concurrently optimize day-ahead energy and reserve needs \cite{Ratha2019}. 

To overcome computational complexity and scalability issues, the usage of distributed optimization for the solution of LEM formulations is suitable and further allows the integration of the relevant cryptographic techniques. The work in \cite{Tian2021} decomposes the collaborative optimization problem using Lagrangian relaxation and solves the master problem via Secure Multi-Party Computation (SMPC) with secret sharing protocols. However, it mainly serves as a proof of concept, with intentionally simple network and market models, and the authors themselves highlight the Lagrangian Relaxation's inability to provide convergence guarantees. Such guarantees can be provided using the Consensus version of the Alternating Direction Method of Multipliers (ADMM) \cite{Boyd2010}, which has already been used with the LinDistFlow grid model \cite{Molzahn2017, Oh2022}.

The main cryptographic technique employed within this study is SMPC, which belongs to a family of cryptographic protocols that can execute arbitrary multiparty computations with information-theoretic security in a threshold model \cite{Maurer2006}. This characteristic renders SMPC well-suited for use in LEM design. Notably, it sidesteps the challenges that previous privacy-preserving optimization methods encountered, such as poor computational performance, degraded convergence guarantees, and weaker security notions like differential privacy \cite{Mak2020, Dvorkin2020}. Furthermore, the applicability of SMPC extends to reinforcing the security of LEMs protecting against network sensor failures due to potential interference of adversaries \cite{Deng2017}. 

This paper proposes a fully privacy-preserving LEM framework based on distributed optimization and SMPC with secret sharing protocols. Building upon prior research \cite{Mieth2019}, we develop an uncertainty-aware joint market for energy and reserves using the chance-constrained LinDistFlow formulation. The model is further extended by introducing batteries as local energy storage \cite{Pozo2022} and by capturing the tariff-switching behavior of the substation \cite{Gerwin2020}. To overcome the computational limitations encountered in \cite{Tian2021}, we propose a distributed and privacy-preserving market clearing mechanism using an SMPC-secured Consensus ADMM algorithm, as described in Sec.~\ref{distopt} and Sec.~\ref{smpc}. The secure version of the LEM requires mechanisms to preserve the security of the system between SMPC sessions, for which we use standard techniques called commitment schemes and Zero-Knowledge Proofs that can be applied without impinging on the overall privacy preservation \cite{Sui2020}. In particular, we leverage SMPC and ElGamal commitments \cite{Kamm2007} to provide a double verification scheme and a measurement recovery scheme (Sec.~\ref{measure}) to provide a secure and verifiable settlement process (Sec.~\ref{dvs}). This setup allows the proposed market mechanism to achieve similar solutions as a central insecure solver but with added privacy preservation, as shown by the results in Sec.~\ref{section:4}.

\section{Local Electricity Market Design} \label{section:2}

\subsection{Preliminaries} 

This paper studies a LEM in an active distribution grid. The distribution network is represented as an undirected and connected tree graph $\mathcal{G}(\mathcal{N},\mathcal{E})$, where $\mathcal{N}=\{0,1,\dots,N\}$ is the set of nodes in the graph and \(\mathcal{E}\subset\mathcal{N}\times\mathcal{N}\) is the set of $N$ edges. The substation node is indexed by 0 and represents the interface between the distribution network and the upper-level grid. Furthermore, we use $\mathcal{N}^+=\mathcal{N}\setminus\{0\}$ to denote the set of non-substation nodes. For a node $n\in\mathcal{N}$, its ancestor node is denoted by $A_n$, while the set of its children nodes is denoted by $\mathcal{C}_n$. The notation related to the physical quantities of the LinDistFlow network model is introduced in the following.

For each node $n\in\mathcal{N}^+$, let $u_n$ denote the squared voltage magnitude at the node, and ${u}_{n}^\mathrm{max}$ and ${u}_{n}^\mathrm{min}$ the corresponding maximum and minimum voltage limits. The substation node is assumed to have a fixed predefined voltage magnitude $u_0$. The active and reactive line flows to a node $n\in\mathcal{N}^+$ from its ancestor node \(A_n\)  are denoted by $f_n^{P}$ and $f_n^{Q}$, respectively. Furthermore, each such line is characterized by a resistance $r_n$, a reactance $x_n$, and a maximum apparent power limit ${S}_n$. 

The participants in the considered LEM include: (i) consumers at all non-substation nodes \(i\in\mathcal{N}^+\), who each have known and inflexible active \(d_i^{P}\) and reactive \(d_i^{Q}\) demand profiles, (ii) prosumers \(r \in \mathcal{R}\subseteq\mathcal{N}^+\) who own DERs with a forecasted active power production \(h_r^f\), (iii) batteries $m \in \mathcal{M}\subseteq\mathcal{N}^+$, characterized by their State of Charge (SoC) \({B}_m\), and the corresponding maximum ${B}_{m}^\mathrm{max}$ and minimum ${B}_{m}^\mathrm{min}$ SOC limits, 
and (iv) the substation node, which is centrally operated to trade energy and procure reserves with the wholesale markets.  Batteries and prosumers constitute flexible generators \(v \in \mathcal{V}=\mathcal{R}\cup\mathcal{M}\) with adjustable active power generation \(g_v^{P}\) between maximum $P_v^\mathrm{max}$ and minimum $P_v^\mathrm{min}$ limits. The cost of active power adjustment is characterized by a quadratic $c^q_v$ and a linear $c_v^l$ cost coefficient. 
The substation node is the ``infinite bus'', with its generation modeled as the difference between the active $l^{P}$ and reactive $l^{Q}$ inflow and active $s^P$ and reactive $s^Q$ outflow to capture a tariff-switching scheme without the need for binary variables \cite{Gerwin2020}, i.e., the system operator charges inflow to the LEM by the forecasted wholesale price plus a flat usage tariff \(\Phi^+\), and pays for outflow at the wholesale price minus the tariff \(\Phi^-\). 

\subsection{Uncertainty Modeling}
The DER generation of a prosumer \(r\in\mathcal{R}\) introduces uncertainty in the LEM and is thus modeled by 
\begin{equation}
    {h}_r=h_r^f + {\omega}_r,
\end{equation} 
where ${h}_r$ is the stochastic DER generation and \({\omega}_r\) is the random forecast error. The forecast error is modeled as an independent Gaussian random variable with zero mean $\mathbb{E}[{\omega}_r]=0$ and variance $\mathrm{Var}[{\omega}_r]=\sigma_r$, known only to the prosumer $r\in\mathcal{R}$ itself.  
The total DER forecast error in the system can thus be calculated as ${\Delta}=\sum_{r\in\mathcal{R}}{\omega}_r$, resulting in a zero-mean multivariate distribution with $\Sigma=\mathrm{Var}[{\Delta}]=\mathrm{diag}(\{\sigma_r\}_{r\in\mathcal{R}})$.

The realized total forecast error creates a power imbalance in the system that requires a global response from the flexible assets. To this end, we equip each flexible generator $v\in\mathcal{V}$ with a linear reserve policy \cite{Ratha2019}, characterized by a participation factor $\alpha_v$, such that its total generation is given by  
\begin{equation}\label{eq:flex_prov}
    \tilde{g}_v^{P} = g_v^{P} - \alpha_v  {\Delta}.
\end{equation}
Given that $\sum_{v\in\mathcal{V}}\alpha_v=1$, the flexible generators completely balance the observed overall active power deviation.

\subsection{Market Clearing Formulation}
In this section, we present the chance-constrained market formulation.
The variable set includes the previously defined network and adjustable generation quantities and is defined by
\begin{alignat}{1}    
\Xi^{\mathbb{P}} =&\, \{u_{i,t},f^P_{i,t}, f^Q_{i,t}\}_{i \in \mathcal{N}^+,t\in\mathcal{T}}\cup \{g_{v,t}^P, \alpha_{v,t}\}_{v\in\mathcal{V},t\in\mathcal{T}}\nonumber\\
&\cup \{l^P_t, s^P_t, l^Q_t, s^Q_t\}_{t\in\mathcal{T}}\cup \{B_{m,t}\}_{m\in\mathcal{M},t\in\mathcal{T}},
\end{alignat}
where $\mathcal{T}$ is the set of considered time steps. 
The aim of the proposed local market is to minimize the expected cost of generation and energy procurement from the upper-level grid while at the same time respecting the network and generation constraints for all time steps $t\in\mathcal{T}$:
\begin{subequations} \label{full_cc_form}
\begin{equation}
    \min_{\Xi^{\mathbb{P}}} \, \mathbb{E}\Big[\sum_{t \in \mathcal{T}}\sum_{v\in\mathcal{V}}\Big(c_{v}^q(\tilde{g}_{v,t}^P)^2 + c_r^l\tilde{g}_{v,t}^P\Big)+ {l}^P_t\Phi^+_t - {s}_t^P \Phi^-_t\Big]  \label{full_cc_form_obj}
\end{equation}
\begin{alignat}{2}
&\text{s.t.}\quad l_t^P - s_t^P = \sum_{j\in\mathcal{C}_0}f_{j,t}^P     &&\label{eq:full_cc_form_1}\\
&l_t^Q - s_t^Q = \sum_{j \in \mathcal{C}_0}f_{j,t}^Q   &&\label{eq:full_cc_form_2}\\
&l^P_t\geq0, s^P_t\geq0, l^Q_t\geq0, s^Q_t\geq 0 &&\label{eq:full_cc_form_17}\\
&f_{n,t}^P + (\tilde{g}_{n,t}^P - d_{n,t}^P + h_{n,t}) = \sum_{j \in \mathcal{C}_n}f_{j,t}^P, &&\forall n \in \mathcal{N}^+ \label{eq:full_cc_form_3}\\
&f_{n,t}^Q - d_{n,t}^Q = \sum_{j \in \mathcal{C}_n}f_{j,t}^Q, &&\forall n \in \mathcal{N}^+ \label{full_cc_form_4}\\
&u_{n,t} = u_{A_n, t} -2(r_nf_{n,t}^P + x_nf_{n,t}^Q), &&\forall n \in \mathcal{N}^+\label{eq:full_cc_form_5}\\
&B_{m,t} = B_{m,t-1} - \tilde{g}_{m,t}^P, &&\forall m\in\mathcal{M} \label{eq:full_cc_form_12}\\
&\mathbb{P}[{u}_{n,t}\leq {u}_{n}^\mathrm{max}]\geq 1-\epsilon_u, &&\forall n\in\mathcal{N}^+ \label{eq:u_max_cc}\\
&\mathbb{P}[{u}_{n}^\mathrm{min}\leq {u}_{n,t}]\geq 1-\epsilon_u, &&\forall n\in\mathcal{N}^+ \label{eq:u_min_cc}\\
&\mathbb{P}[a^1_n{f}_{n,t}^P + a^2_n{f}_{n,t}^Q + a^3_n S_{n}\leq 0] \geq 1-\epsilon_f,\quad &&\forall n \in \mathcal{N}^+\label{eq:line_cc}\\
&\mathbb{P}[\tilde{g}_{v,t}^P\leq P_v^\mathrm{max}]\geq 1-\epsilon_g, &&\forall v\in\mathcal{V} \label{eq:g_max_cc}\\
&\mathbb{P}[P_v^\mathrm{min}\leq\tilde{g}_{v,t}^P] \geq 1-\epsilon_g, &&\forall v\in\mathcal{V} \label{eq:g_min_cc}\\
&\mathbb{P}[B_{m,t}\leq B_m^\mathrm{max}]\geq 1-\epsilon_b, &&\forall m\in\mathcal{M} \label{eq:bat_max_cc}\\
&\mathbb{P}[B_m^\mathrm{min}\leq B_{m,t}]\geq 1-\epsilon_b, &&\forall m\in\mathcal{M} \label{eq:bat_min_cc}\\
&\sum_{v\in\mathcal{V}}\alpha_{v,t}=1,\quad 0\leq\alpha_{v,t}\leq 1, &&\forall v\in\mathcal{V} \label{eq:part_fac_c}.
\end{alignat}
\end{subequations}
Constraints related to inflow/outflow at the substation node are given in \eqref{eq:full_cc_form_1}-\eqref{eq:full_cc_form_17}. The LinDistFlow network model is established in \eqref{eq:full_cc_form_3}-\eqref{eq:full_cc_form_5}, and the battery SoC model\footnote{For the sake of simplicity, but without loss of generality, we assume perfect charging and discharging efficiencies.} in \eqref{eq:full_cc_form_12}. The chance constraints on bus voltages are enforced in \eqref{eq:u_max_cc}-\eqref{eq:u_min_cc}, power generation limits in \eqref{eq:g_max_cc}-\eqref{eq:g_min_cc}, battery SoC in \eqref{eq:bat_max_cc}-\eqref{eq:bat_min_cc}, and the dodecagon linear approximations (defined by coefficients $a^1_n,a^2_n,a^3_n,\forall n\in\mathcal{N}^+$) of the line flow constraints in \eqref{eq:line_cc}. The LEM operator specifies an error term for each type of CC, representing the maximum acceptable percentage of a constraint violation, namely, \(\epsilon_g\) for generation limits, \(\epsilon_b\) for battery limits, \(\epsilon_u\) for voltage limits and  \(\epsilon_f\) for flow limits. Finally, the bounds on reserve coefficients are given in \eqref{eq:part_fac_c}. 

\subsection{Distributed Solution Method}\label{distopt}
To obtain a tractable form of the optimization problem in \eqref{full_cc_form}, each individual linear chance constraint can be reformulated as a second-order cone constraint. This process is omitted for brevity, and we refer the reader to \cite{Mieth2019} for more details. Instead, we here focus on decomposing the centralized optimization problem into a distributed form using the scaled formulation of Consensus ADMM \cite{Boyd2010}, as proposed in \cite{Oh2022}.

Considering that the objective function in \eqref{full_cc_form_obj} is separable, let us introduce a cost function $f_n(X_n)$ related to each node $n\in\mathcal{N}$, where $X_n$ denotes the vector of all variables related to node $n$. A subset of these variables, collected in $X_{\mathbb{C}_n}=(u_n, f_n^P,f_n^Q,\alpha_n)$ and called coupling variables, appear in the constraints of other nodes. The coupling variables have a global copy $Z_{\mathbb{C}_n}$ in the Consensus ADMM algorithm that ensures their system-wide convergence. The full algorithm at iteration $k$ is given by
\begin{subequations}\label{eq:admm_update}
\begin{align}
X_{\mathbb{C}_n}^{k+1} &= \mathrm{arg\,min}\,{f}_n(X_n)+\frac{\rho_0}{2} ||X_{\mathbb{C}_n}^k - Z_{\mathbb{C}_n}^k+ U_{\mathbb{C}_n}^k||_2^2, \\
Z^{k+1}_{\mathbb{C}_n} &= \mathbb{AVG}(X^{k+1}), \label{eq:z_avg} \\ 
U_{\mathbb{C}_n}^{k+1} &= U_{\mathbb{C}_n}^{k} + X_{\mathbb{C}_n}^{k+1} - Z_{\mathbb{C}_n}^{k+1},   
\end{align} 
\end{subequations}
where $\rho_0$ is a positive scaling factor and $U_{\mathbb{C}_n}$ are the Lagrange multipliers. In \eqref{eq:z_avg}, an arithmetic mean of all relevant local values is computed to find the global values of the coupled variables.
The employed convergence metrics are the $\ell_2$-norms of the primal and dual residuals of the local problems and the total power surplus in the system relative to the total demand.

\subsection{Financial Settlement}\label{sec:settle_theory}
Solving the market problem \eqref{full_cc_form} yields nodal \(\lambda_n\) and flexibility $\pi$ prices, calculated as dual variables of constraints \eqref{eq:full_cc_form_3} and \eqref{eq:part_fac_c}, respectively. Using these prices, the financial payoff for a market participant \(n\) from the daily market operation is 
\begin{equation}\label{eq:finance}
    \mathcal{B}_n = \sum_{t \in \mathcal{T}}\Big[\lambda_{n,t}^P(g_{n, t}^P + h_{n, t}^f - d_{n, t}^P) + \pi_t(\alpha_{n, t} - \frac{\sigma_{n, t}}{\sum_{r \in \mathcal{R}}\sigma_{r, t}})\Big]
\end{equation}
As can be seen above, the participants are charged at the flexibility price \(\pi\) for their share of the overall uncertainty, captured by the standard deviation of their forecast error, as well as a fixed tariff for the power flow via the substation \cite{Gerwin2020}. Their profit stems from the provided flexibility and the injected DER controllable and forecasted active power.
The final balance \(\mathcal{F}_n\) of a party \(n\), however, also includes the results from the two-price imbalance regulation \(\mathcal{I}_n\) performed by the DSO \cite{Jonsson2014} on any remaining deviations:
\begin{equation}\label{eq:final_bal}
    \mathcal{F}_n = \mathcal{B}_n + \mathcal{I}_n.
\end{equation}

\section{Secure Local Market Protocol} \label{section:3}
\subsection{Secure Multiparty Computation Preliminaries}
The technique used to ensure the security of the market protocol is Secure Multiparty Computation \cite{Maurer2006}, which allows a group of parties to execute secure computations without relying on a trusted third party. The communication links between the parties are assumed to be encrypted and synchronous. Security in SMPC is defined by input privacy and protocol correctness. Input privacy requires the input values and calculation results to be kept hidden from other participants unless intentionally revealed, whereas protocol correctness requires that the computations yield the same results as with a trusted third party. The goal is thus to have security no worse than an ideal scheme with a trusted party, i.e., SMPC is allowed to be vulnerable to attacks that also work against the ideal scheme. 

The considered adversaries are assumed to be static and honest-but-curious. The static adversary corrupts participants before the start of the protocol but does not corrupt further during execution. Secondly, the honest-but-curious (or passive) adversary seeks only to violate the input privacy property and does otherwise not deviate from the established SMPC protocol \cite{Maurer2006}. The choice to work with an honest-but-curious adversary is due to the nature of ADMM. While the global stage (iteration) can be implemented with an SMPC secure against misbehaving or active adversaries, these techniques cannot protect against parties misbehaving in the local optimization stage. In particular, an adversary could simply return random values at each iteration and thereby prevent the ADMM from converging without violating any of the SMPC guarantees. Overcoming this issue would necessitate delving into verifiable computing aspects, which are out of the scope.

\subsection{Shamir Secret-sharing Scheme}
The SMPC protocol used in this paper is based on the Shamir secret-sharing scheme (SSS) \cite{Tian2021}.  It provides threshold security based on the number of participants $N$ and a defined threshold \(\Theta<N\), where a single, central adversary can corrupt up to $\Theta$ participants without compromising the protocol's security. 
From a high-level perspective, the scheme involves transforming a secret $s$ into a secure \textit{shared} domain, yielding $[s]$, then performing secure calculations on the shared values, and finally, recovering relevant results $r$, by a reconstruction procedure $[r] \mapsto r$. It is worth noting that due to the inner workings of Shamir secret sharing \cite{Tian2021}, calculations in an SMPC scheme based on it can only use addition and multiplication operations. As such, the secure market protocol that follows uses a variety of transformations and simplifications to reduce computational complexity as much as possible. 

\subsection{Secure Market Protocol Overview}
\begin{figure}[!b]
\includegraphics[width=1\linewidth]{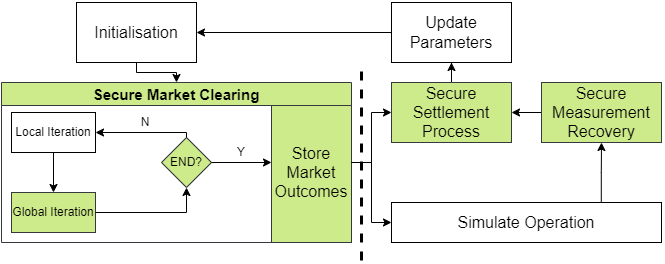}
\caption{Overview of the complete (secure) market protocol.}
\label{fig:protocol}
\end{figure}
The overview of the proposed market protocol is presented in Fig.~\ref{fig:protocol}, with the secure blocks given in green and a dashed line separating the time of market clearing and the time of operation.
Each day, market participants initialize the secure market clearing protocol by providing their desired internal parameters such as battery limits, cost coefficients, generation limits, etc. The secure market clearing is then performed, as will be explained in Sec.~\ref{smpc}, which involves the distributed market formulation from Sec.~\ref{distopt} with the updating of global variables and testing for convergence, both implemented securely with SMPC. Upon completion, the parties then store the outcomes of the market for later usage. After the simulated or real market operation, the parties then execute a secure measurement recovery routine (Sec~\ref{measure}), which ensures the protocol has all the relevant information it needs to proceed. Then, the secure settlement process (Sec.~\ref{dvs}) uses the stored market outcomes and recovered measurements to securely compute the financial balances of all parties. The protocol then ends for the day by having the parties update their internal parameters based on the financial and operational outcomes.

\subsection{Secure Market Clearing}\label{smpc}
\begin{figure}[t!]\centering
\includegraphics[width=1\linewidth]{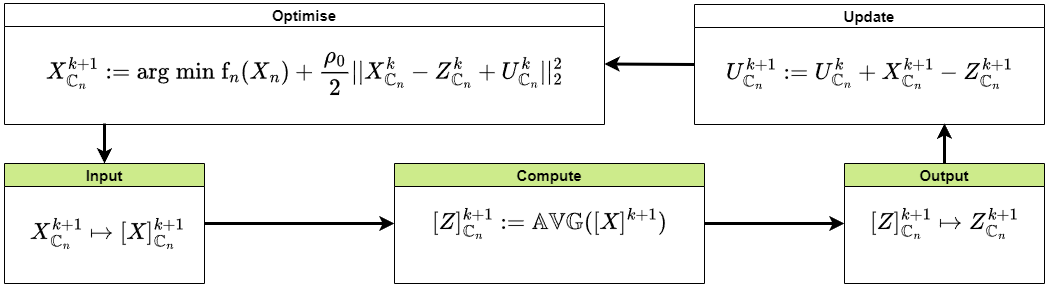}
\caption{Illustration of an ADMM iteration with SMPC.}
\label{fig:sec_admm}
\end{figure}

The main usage of SMPC is to secure the distributed optimization process by replacing the central node that performs the coordination calculations \eqref{eq:z_avg}. Figure \ref{fig:sec_admm} shows how the ADMM update loop from \eqref{eq:admm_update} can be secured via SMPC, where white boxes are local operations and boxes shaded green are SMPC operations. Parties hereby securely share the new value of their local variables $X^{k+1}_{\mathbb{C}_n}$, then calculate the shared value of the new global variables $[Z]_\mathbb{C}^{k+1}$, and finally, output their true values to the relevant parties for use in local optimization $Z^{k+1}_{\mathbb{C}_n}$. To improve performance, the division operation in the algebraic mean calculation is done locally, meaning the SMPC instance only involves the addition of scaled variables. 

The evaluation of convergence criteria is also done within the SMPC instance, which can become computationally expensive due to the non-linear complexity of secure inequality evaluations and Euclidean norm calculations. To address this issue, a stricter version of the residual criteria is used -- the infinity norm, instead of the commonly employed $\ell_2$-norm.

\subsection{Measurement Recovery}\label{measure}
A major concern in digital markets, including the considered LEM, is the ``Oracle Problem", where a cyber attack on the underlying measurement systems cause the ground truth and the market's understanding to diverge, causing strain on consumers and network operators \cite{Caldarelli2020}. This problem is exacerbated in distribution grids where ensuring complete network observability would require expensive deployment of PMUs or RTUs  \cite{Dehghanpour2019, Wang2019}. The market protocol is thus backstopped with a deterministic and secure measurement recovery procedure to ensure the financial settlement process can be executed even if a subset of nodes fails to report any measurements. 

The two main failure modes are nodes not reporting measurements and nodes reporting false measurements \cite{Deng2017}, with the latter being difficult to counteract if done at scale. This paper focuses on the former, with the use case being participants purposefully disconnecting their deployed sensor during operation. We assume the use of modified smart meters that can measure the active and reactive nodal injections and the active and reactive incoming line flow at each node. The measurements along the lines are not required since LinDistFlow assumes lossless lines. Under the assumption of altered smart meters and an always reporting substation node, the following recovery procedure generates feasible values for all outstanding net injection measurements, even if they are not guaranteed to report the actual value. 

Non-reporting nodes in the network are grouped into disjoint segments, called \textit{islands}, via the UnionFind algorithm \cite{Galli1991}. For each such island \(\mathcal{Y}\), SMPC is then used to calculate its total active power inflow \(\mathrm{P}_{\mathcal{Y}}^\mathrm{in}\) and outflow \(\mathrm{P}_{\mathcal{Y}}^\mathrm{out}\) using the measurements of honest ancestors and descendants. This determines the overall active power net injection of the island, which is then split equally among the members of a given island to ensure burden sharing, as follows
\begin{equation}\label{eq:measure_recov}
\mathrm{net}_{y}^{P} = \frac{\mathrm{P}_{\mathcal{Y}}^\mathrm{out} - \mathrm{P}_{\mathcal{Y}}^\mathrm{in}}{|\mathcal{Y}|},\quad\quad\quad\forall y\in\mathcal{Y},
\end{equation}
where $|\mathcal{Y}|$ denotes the cardinality of set $\mathcal{Y}$.
Note that the resulting value may not correspond to the withheld value for each $y\in\mathcal{Y}$. Any such deviations are balanced via the balancing mechanism at the time of operation and factor into the settlement process.

\subsection{Secure Settlement Process}\label{dvs}
The calculation of the financial payoff \(\mathcal{B}_n\) for a party \(n\) after the market clearing is performed employing the same SMPC protocol used for ADMM coordination. Since the prices used for calculating \(\mathcal{B}_n\) are dual variables of constraints, they can also be calculated globally as Lagrange multipliers via SMPC, as explained in \cite{Zhang2019}. For the final financial balances, the settlement process then factors in the results of the DSO imbalance market (see Section \ref{sec:settle_theory}) and the measurement recovery process.

The secure market protocol introduces a break in time between market clearing and final settlement, requiring each node to re-input their own balance \(\mathcal{B}_n\) into the SMPC instance computing the final settlement. A Double Verification Scheme (DVS) is therefore proposed to allow secure verification of these input balances by an honest majority of parties, including actors not participating in the protocol, such as the market operator. The DVS combines storing and securely comparing secret-shared balances with the use of cryptographic commitments to verify the self-claimed balances and is inspired by publicly verifiable secret-sharing schemes \cite{Schoenmakers1999}. Specifically, parties re-share their post-market clearing financial balances, which are securely compared using SMPC to the shared values the other parties stored after market clearing. In the second phase, parties use the perfectly binding ElGamal commitment scheme to ``commit" to their balances after market clearing. If needed, they can then ``open" the balances in the second phase of DVS to prove that they used the correct value without it actually being revealed \cite{Kamm2007}. Thus, DVS Phase 1 provides threshold security via SMPC and DVS Phase 2 provides information theoretic security via ElGamal commitments.

DVS thus protects against manipulations by corrupted parties while also providing cryptographic proof for the calculated final financial balances. The secure settlement can then use SMPC to calculate imbalances between the scheduled and actual network values that have both been input securely to find \(\mathcal{I}_n\) for each party and then output the final financial balances of all parties. 

\section{Results}\label{section:4}

\subsection{Case Study Setup}
\begin{figure}[!b]
\centering
\includegraphics[width=0.99\linewidth]{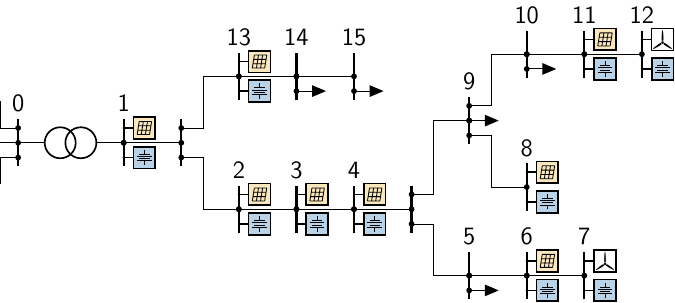}
\caption{A synthetic 15-bus test system, with PV generation (in yellow), batteries (in blue), and wind generation (in white).}
\label{fig:simnet1}
\end{figure}

The proposed secure market clearing protocol is evaluated on a 15-bus test network depicted in Fig.~\ref{fig:simnet1}. This synthetic case study is constructed using simulated data from various real-world datasets, as described in the following. The network model incorporates nodal demand and solar power generation profiles from an English dataset \cite{CLNR}, wind power generation profiles from an Australian dataset \cite{Dowell2015}, and battery sizing based on the Tesla Powerwall \cite{Moore2023}. Note that in this specific case study, all prosumers have both a DER and a battery, with their flexible generation stemming entirely from the battery. The power prices are obtained from various sources, including suggested substation tariff \cite{Sousa2019}, wholesale day-ahead prices \cite{Energinet2022a}, and regulation market prices \cite{Energinet2022b} from the Danish TSO. Finally, the cost parameters and standard deviations for the prosumers' generation are based on the work in \cite{Mieth2019}, and the battery parameters are derived from \cite{Castano2015}.

The Python package MPyC \cite{Schoenmakers2018} was utilized for implementing the SMPC protocol and additionally provided the elliptic curve cryptography necessary for the ElGamal commitment scheme. For the local optimization, the CVXPY optimization library was used with ECOS solver for second-order cone programs and the OCSP solver for quadratic programs. CVXPY also natively supports ADMM. The implementation is done in Python, and the code was executed on a laptop with a 6-core processor and 16 GB of RAM. 

The results section centers around a comparative analysis of the performance of the following four local market solvers:
\begin{itemize}
    \item \textbf{C-1}: An insecure central solver solving the market problem in \eqref{full_cc_form}, with all chance constraints reformulated as second-order cone constraints \cite{Mieth2019}.
    \item \textbf{N-3}: An insecure distributed solver solving the reformulated chance-constrained market problem \eqref{full_cc_form}.
    \item \textbf{S-2}: A secure distributed solver solving the reformulated chance-constrained market problem \eqref{full_cc_form} with deterministic voltage magnitude and line flow constraints. Security is imposed using the protocols described in Sec.~\ref{section:3}.
    \item \textbf{S-3}: A secure distributed solver solving the complete reformulated chance-constrained market problem \eqref{full_cc_form}, making it the secure version of N-3.
\end{itemize}
\subsection{Solver Comparison}
The level of security offered by these solvers depends on the extent to which they disclose the private data of the parties during operation. The central solver (C-1), for instance, discloses all parties' information solely to a central computation node. If this central point is compromised, it could potentially threaten input privacy and the correctness of the entire market protocol. On the other hand, in the distributed solvers utilizing ADMM, nodes share certain data only with their immediate neighbors (N-3). Furthermore, when secured with SMPC, widespread data leakage can only occur if a majority of parties are compromised (S-2 and S-3). 

Table~\ref{fig:baseline} compares the four solvers in terms of the time and global iterations required for the algorithm to converge, as well as the resulting relative accuracy. The latter refers to the total active power residual in the LEM relative to the fixed total demand and gauges the feasibility of the solution. All considered solvers converge with similar levels of relative accuracy, but the use of both the decomposition and SMPC extend the time required for the convergence. Notably, N-3 and S-3 require similarly many ADMM iterations, indicating that the calculations in SMPC work adequately. However, comparing S-2 and S-3 shows that full chance constraining doubled the ADMM iterations and tripled the time to convergence. This points to a trade-off between the model complexity and the computation performance. Overall, the results indicate that the secure market protocol via solvers S-2 and S-3 is capable of clearing a complex market with results that are slightly worse but comparable to the centralized solver.

\begin{table}[!b]
\vspace{-0.35cm}
\renewcommand{\arraystretch}{1.25}
\caption{Comparison of the four featured solvers.}
\label{fig:baseline}
\noindent
\centering
    \begin{minipage}{\linewidth}
    \renewcommand\footnoterule{\vspace*{-5pt}} 
    \begin{center}
        \begin{tabular}{ c || c | c | c  c }
            \toprule
            \textbf{Solver} & \textbf{C-1} & \textbf{N-3}  & \textbf{S-2} & \textbf{S-3} \\ 
            \cline{1-5}
            Time [min] & 0.49 & 66.76 & 38.06 & 119.8 \\
            Global iterations & 1 & 595 & 270 & 593\\
            Rel. accuracy [\%]& $0$ & $1.8$ & $-1.96$ & $4.21$\\
            \bottomrule
        \end{tabular}
        \end{center}
    \end{minipage}
\end{table}

\subsection{Convergence of Secure Market Protocols}
\begin{figure}[!t]
\centering
\includegraphics[scale=1]{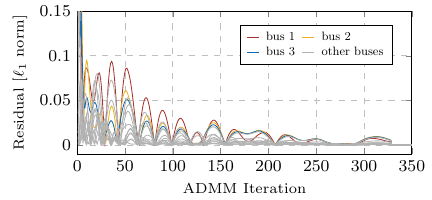}
\caption{Voltage magnitude residuals across all nodes when using S-2 solver.}
\label{fig:u_res}
\vspace{-0.35cm}
\end{figure}

As a result of the use of ADMM, all distributed solvers in the paper are guaranteed to eventually converge \cite{Zhang2019}. However, as the convergence of the residuals of the voltage magnitudes for the S-2 solver in Fig.~\ref{fig:u_res} highlights, there are some notable dynamics at play. The convergence of ADMM typically exhibits a damped oscillatory behavior \cite{Boyd2010}, as is also evident in the given figure, persisting until approximately iteration 250. Around this iteration, the residuals start to briefly increase again due to the inter-dependency of voltage magnitudes and power flows in the LinDistFlow model. Their constant trade-off (as well as the ripple effect across the nodes) implies that the solver can locally diverge from the expected trends but will eventually recover as the ADMM iterations progress. Interestingly, the residual oscillations are of higher magnitudes at nodes closer to the substation. The observed convergence pattern highlights the possibility for improvement, e.g., through the implementation of adaptive penalty terms or over-relaxation techniques \cite{Zhang2019}, which are, however, beyond the scope of this paper. 

\subsection{Market Clearing Outcomes}
\begin{figure}[b]
\vspace{-0.35cm}
\centering
\includegraphics[scale=1]{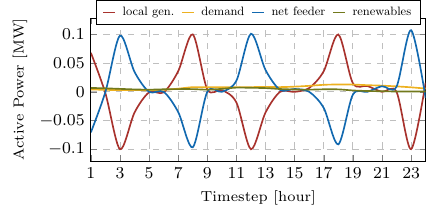}
\caption{Global energy balance over a day for the S-2 solver.}
\label{fig:schedules2}
\end{figure}
Figure \ref{fig:schedules2} showcases the different energy balance components over the entire network. It can be observed that the day-ahead energy dispatch of the network is driven by prosumers who use batteries to shift the DER production into the high-demand morning and evening hours. Similar behavior has been seen in previous research \cite{Zhang2020}. Despite the slight changes in formulations and solution methods, all solvers had very similar energy schedules and prices as the presented S-2 solver, indicating that the secure market protocol is able to clear the market successfully.

The allocation of flexibility participation factors described in \eqref{eq:flex_prov} by the different solvers is observed to fall into one of the following two arrangements, which is then maintained for the entire day.
Specifically, all fully chance-constrained solvers assigned the substation the entire flexibility provision, while the S-2 solver, which does not factor in voltage and line flow constraints, splits the flexibility provision equally amongst all nodes with flexible generation. This resulted in the flexibility price being constant throughout the day.

\subsection{Financial Settlement Results}
The above-discussed scheduling outcomes are reflected in the post-market clearing financial balances, with Table \ref{fig:fin_balances} comparing the balances of 4 different LEM participants between solvers N-3, S-2, and S-3. Due to the large variability found in flexibility prices across solvers, the table focuses on the day-ahead energy results. As explained in Sec.~\ref{dvs}, in a distributed setting, the market prices can either originate from the dual variables in the local optimization problems that the parties then share with the LEM (Duals) or from global Lagrange multipliers calculated in SMPC (Sec-Bal). To evaluate the impact of using SMPC on these critical calculations, both secure solvers are considered with both price origins and compared to the insecure, distributed solver N-3. The nodes selected for evaluation are the substation node 0, the PV-prosumer node 3, the wind-prosumer node 7, and the pure load at node 15. Across solvers and price origins, there are only minor differences arising from the usage of SMPC, highlighting both SMPC's viability in the LEM and the overall scheme's success at secure market operation.  

\begin{table}[!t]\label{table:bal}
\renewcommand{\arraystretch}{1.2}
\caption{Final Day-Ahead Energy Balances in CHF at several nodes for Solvers N-3, S-2, and S-3.}
\label{fig:fin_balances}
\noindent
\centering
    \begin{minipage}{\linewidth}
    \renewcommand\footnoterule{\vspace*{-5pt}} 
    \begin{center}
    \scalebox{0.81}{%
        \begin{tabular}{ c || c | c  c | c c }
            \toprule
            \textbf{Solver} & \textbf{N-3} (Duals) & \textbf{S-2} (Duals)  & \textbf{S-2} (Sec-Bal) & \textbf{S-3} (Duals) & \textbf{S-3} (Sec-Bal) \\ 
            \cline{1-6}
            Node 0 & $-8.76$ & $-7.81$ & $-7.83$ & $-8.41$ & $-8.40$ \\
            Node 3 & $0.47$ & $0.43$ & $0.41$ & $0.47$ & $0.47$\\
            Node 7 & $3.71$ & $3.84$ & $3.84$ & $3.71$ & $3.71$\\
            Node 15 & $-0.88$ & $-0.92$ & $-0.92$ & $-0.88$ & $-0.88$\\
            \bottomrule
        \end{tabular}
        }
        \end{center}
    \end{minipage}
    \vspace{-0.35cm}
\end{table}

\section{Conclusion}\label{section:5}
In this paper, we presented a Local Electricity Market framework in a distribution network with uncertain distributed energy resources. To preserve the input privacy of the market participants, we solved the market problem using an ADMM-based distributed optimization with data exchanges protected by leveraging secure multi-party computation protocols. Amongst the considered secure solvers, S-2 stands out as the recommended choice. It avoids the single point of failure of C-1, has security guarantees via SMPC that N-2 lacks, and achieves comparable results in fewer iterations and minutes than S-3. Note that since the secure market protocol is highly customizable, e.g., in its convergence thresholds, chance constraint bounds, etc., the choice of the preferred solver may vary. Additionally, the protocol may benefit from an increased parallelization or the use of better hardware, which could make S-3 become more suitable.

\section{Acknowledgement}
This research was supported by NCCR Automation, a National Centre of Competence in Research, funded by the Swiss National Science Foundation (grant number 51NF40\_180545).

\bibliographystyle{IEEEtran}
\bibliography{mybib}

\end{document}